\documentclass[12pt]{article}
\usepackage{amsmath}
\begin{document}
\bibliographystyle{plain}
\title{ Quantum Theory within the Framework\\
of General Relativity. A Local Approach.}
 
\author{P.~Hrask\'{o}\footnote{email: hraskop@freemail.hu}\\
Department of Theoretical Physics,\\ Janus Pannonius
University, Ifjus\'{a}g u.\ 6, 7624 P\'{e}cs, Hungary}
\maketitle
 
\begin{abstract}
A local conception (in the sense of the equivalence
principle) is proposed to reconcile
quantum theory with general relativity, which allows one to avoid
some difficulties --- as e.g. vacuum catastrophe --- of the global
approach. All nonlocal aspects of quantum theory, including
EPR paradox, remain intact.
\end{abstract}
 
\section{Introduction}
 
Quantum theory on a given space-time has been the subject of
intensive study over the past decades \cite{Birrel82} . The
announcement by
S.W.Hawking of the thermal radiation of black
holes \cite{Hawking74} and the
analysis of the behaviour of accelerated radiation detectors by
W.G.Unruh \cite{Unruh76} are outstanding instances of this
research.
 
The quantization procedure adopted in these works was based on the
positive and negative frequency modes extended over
Cauchy surfaces across space-time as a whole. Due to this basic
feature this approach may be characterized as {\em global}. The
greatest difficulty of this field of research consists in the
complete lack of relevant experimental evidences so that it must
be carried out deductively on purely theoretical grounds.
Therefore,
at present we have no definitive observational facts which could
corroborate or refute the basic principles of the global theory.
 
The purpose of the present work is to call attention to an
alternative way to deal with the same subject which is motivated
mainly by some difficulties of the global approach. The foremost
problem we have in mind is the cosmological constant problem (or
vacuum catastrophe) which has already a long
history \cite{Weinberg88}.It consists
in the enormous vacuum energy of particle quantum fields in the
curved space-time around the Sun which should have led to the
complete distortion of the orbits of the planets. A less known
difficulty is connected with the frequency mixing in geodesic
motion in curved space-time. The term "frequency mixing" expresses
the fact that the negative frequency modes present in the vacuum
of a quantum field acquire positive frequency components as seen
by a moving body. It is this phenomenon which is the origin of the
Unruh-effect: Photon detectors accelerated in the empty
pseudo-Euclidean space-time of special relativity are expected
to be permanently excited while continuously emitting
photons. In a curved space-time as e.g. in that around the Sun
frequency
mixing may occur even for a non-accelerated (freely falling)
detector. Though the rate of this effect is far too small to be
observable, its existence acquires principal importance if
considered in the light of the Einsteinian principle of
equivalence.
 
As it is well known this principle asserts \cite{Will81} that
macroscopic bodies
of reference, falling freely without rotation, are the only true
realizations of inertial frames in Nature. They possess all the
basic properties of the
inertial frames of reference introduced in special relativity with
a single exception --- they are of finite rather than
infinite extension (locality vs. globality). A freely orbiting
spacecraft if it does not rotate is the simplest example of a
body of reference, carrying a local inertial frame of Einsteinian
sense.
If bodies moving freely on geodesics experience indeed
frequency mixing then a photon detector which is at rest in
such a spacecraft will be continuously absorbing and emitting
photons. This is,
however, in sharp contradiction with the principle of equivalence
which requires that physical phenomena in local inertial frames
take place just as in the inertial frames of special relativity.
In particular,
photon detectors at rest in vacuum must remain --- after, perhaps,
some relaxion --- in their ground state.
 
Both the vacuum catastrophe and the frequency mixing in geodesic
motion are connected with the infinite extension of the
modes in space-time on which the global quantization procedure
rests.
The approach to be outlined in the following sections will be {\em
local} in the sense that it requires global modes
characteristic to special relativity which are
insensitive to the curvature of space-time and correspond more
closely to the spirit of the equivalence principle. The term
"local"
does not, therefore, imply any change in the nonlocal features of
quantum theory. It refers to the way quantum theory is
reconciled with the space-time of general relativity.

\section{The two types of coordinates. Freely falling bodies
of reference.}
 
It is a well known fact that Dirac-spinor fields can be defined
only
in rectilinear (pseudo)-orthogonal coordinates \cite{Wald84} since
spinor transformation
rule cannot be generalized from (hyperbolic) rotations to general
coordinate transformations. As a consequence, on curved space-time
spinors can be described only with respect to some orthonormal
tetrad field. In spite of this complication spinors play a highly
important role in quantum physics (and also in general
relativity). This fact leads to the intriguing suspicion
that, perhaps, orthonormal tetrads may be more significant
for quantum theory in curved space-time than just being
an auxiliary device.
 
On the world line $\cal L$ of every massive object an orthonormal
tetrad
is naturally defined by Fermi-Walker transport (see the Appendix)
which for geodesics
reduces to parallel transportation. The time-like element of the
tetrad coincides with the unit tangent vector to the world line.
This tetrad field will serve as the starting point for the
construction of two different types of coordinates.
 
The first of them are the Fermi-coordinates attached to $\cal L$.
To obtain the Fermi coordinate-time $t$ of an event $E$ one has to
drop a perpendicular to $\cal L$ from $E$ and identify $t$ with
the proper
time $\tau$ of the point $P$ of intersection of the two curves.
The
perpendicular $PE$ is a space-like geodesic whose 3-direction at
$P$ with respect to the tetrad at that point is given by the polar
angles $\vartheta$, $\varphi$. Then the Fermi space-coordinates
$x,\:y,\:z$ of $E$ are equal to
\[x=l\cdot\sin\vartheta\cos\varphi ,\qquad
  y=l\cdot\sin\vartheta\sin\varphi ,\qquad
  z=l\cdot\cos\vartheta ,\]
where $l$ is the geodesic distance between $P$ and $E$.
 
In this section we confine ourselves to the case when the body of
reference performs free geodesic motion, i.e. its world line is
the geodesic $\cal G$. The main virtue of the Fermi-coordinates is
that, using them, the metric in this case becomes Minkowskian
($g_{ij}({\cal
G})=\eta_{ij}$) and the connection coefficients vanish
($\Gamma_{jk}^i({\cal G})=0$) {\em along the geodesic $\cal
G$} \cite{Synge60, Schouten54} . Therefore, the Fermi-coordinates
are the generalizations of the Minkowski-coordinates in the global
inertial frames of special relativity to the local inertial frames
of general relativity.
It is the Fermi-coordinates which are naturally employed by the
astronauts on a freely orbiting nonrotating spacecraft, assuming
their abode to be at rest and their clocks to show the time.
Though the relations $g_{ij}=\eta_{ij}$ and $\Gamma_{jk}^i=0$ are
--- strictly speaking --- valid only in a single point within the
spacecraft they remain practically valid in some
neighborhood {\sf N} of it. This neighborhood is in fact a rather
extended one since it is the same domain where tidal forces are
negligibly small. It is this {\sf
N} where our physical theories must be valid in their purest form
established for the global inertial frames of special relativity
in Minkowski-coordinates. In particular, the
experiments in quantum physics which can be
performed within {\sf N} must give results in conformity with the
theoretical predictions made in special relativity.
 
The second type of coordinates will be called {\em parametric}.
They will be assumed numerically equal to the Fermi-coordinates
but the manifold which they parametrize is of pseudo-Euclidean
structure whose metric in parametric coordinates is
Minkowskian.
Parametric coordinates cover the whole of the parametric space
while Fermi-coordinates cover only part of the space-time.
The important point is that in the
neighborhood {\sf N} the metrical properties of the space-time do
not discriminate between the space-time Fermi-coordinates and the
parametric coordinates.
 
We are now prepared to formulate the basic assumption of our local
approach: We assume that {\em the vectors of the quantum
theoretical Hilbert-space are labelled by parametric rather than
space-time coordinates.} In other words the coordinates in the
Schr\"{o}dinger-equation are parametric coordinates. This means
that quantum theory works in parametric space though the
phenomena it describes take place, of course, in space-time.
 
Let us illustrate the interplay of the two types of coordinates by
the calculation and interpretation of the emission of a photon by
an atom which is at rest in the reference body moving on $\cal G$.
 
The calculation of the decay probability is performed in
parametric coordinates in exactly the same way as in special
relativity, because the parametric space is by definition
pseudo-Euclidean and the metric induced in it by the
Fermi-coordinates is Minkowskian. The decay constant given by this
calculation may slightly differ from that which can be obtained
by the space-time quantum electrodynamics of the global
approach since the modes in the latter case reflect the curvature
of the space-time to which the parametric space is insensitive.
Were the difference sufficiently large the comparison of the
calculated decay constants with observation would permit us to
choose between the two approaches. But even if the numerical
disagreement is very small it reflects the principal difference
between the global and local points of view. The space-time modes
in terms of which the global approach is formulated are
"realistic" in the sense that they reflect the curvature of the
far regions of space-time as if they were in some way really
there. The local point of view, on the contrary, assumes that
submicroscopic phenomena, such as the decay of an excited atom,
are truly local effects whose rates are fully determined by local
environment. If this idea is correct it would be a
misinterpretation to assign the modes occurring in these
calculations to space-time. I stress that this inclination toward
local description is suggested --- if not dictated --- by the
principle of equivalence to which, as explained in the
Introduction, the global approach seems to contradict.
 
In parametric space the rays of the emitted light are zero
geodesics. The identification of the parametric coordinates with
the space-time Fermi-coordinates permits us to map these rays into
space-time. But this mapping does not in general conserve the
geodesic nature of world lines. So,
concerning light deflection in gravitational field, we arrive at a
contradiction with the experience
since the explanation of light deflection is based on the fact
that light rays are null geodesics in space-time.
Therefore, the relegation of quantum theory to the parametric
space leads to the unavoidable consequence
that its predictions which are not
strictly local become increasingly worse when moving off the
world line of the body of reference whose parametric space is the
terrain of the calculations.
 
The proper treatment of this phenomenon is made possible by the
fact just mentioned that the photons travelling in space between
distant bodies
of reference follow quasiclassical trajectories, i.e. move on zero
geodesics. In the calculation of their emission they appear as the
out states $\vert k_1,\epsilon_1;out\rangle$ of definite momentum
and polarization. The momentum is defined by the tangent to the
trajectory at its initial point with respect to the
reference frame 1 where the emission takes place. Similarly, the
absorption of the photon on the body 2 takes place from the
in-state $\vert k_2,\epsilon_2;in\rangle$. The "propagation
matrix"
$_2{\langle k_2,\epsilon_2;in\vert k_1,\epsilon_1
;out\rangle}_1$ is determined by the quasiclassical motion of the
photons between the reference bodies 1 and 2 in conformity with
the fact that our knowledge of both light deflection and red shift
which are the paradigms of the phenomena under consideration comes
entirely from the classical theory of light propagation. The role
of quantum theory in these processes is confined to the
determination of the emission rates and absorption cross sections.
It may be noted that the procedure just
described is nothing but the present day practice made into
principle.
 
At the conclusion of this section we add on more argument in favor
of parametric space quantization to those brought forward in the
Introduction.
It is well known that canonical quantization leads to a theory
consistent with experiment only if it is performed in terms of
Cartesian coordinates and momenta. In spherical coordinates, for
example, commutation relations $[\hat\varphi ,\hat p_\varphi ] =
i\hbar$ etc. are easily satisfied by $\hat\varphi =\varphi$ etc.
and momentum operators
\[\hat p_r = \frac{\hbar}{i}\left (\frac{\partial}{\partial
r}-\frac{2}{r}\right ),\qquad
\hat p_\vartheta = \frac{\hbar}{i}\left
(\frac{\partial}{\partial\vartheta}-\cot\vartheta\right ),\qquad
\hat p_\varphi =
\frac{\hbar}{i}\frac{\partial}{\partial\varphi},\]
which are Hermitean (more precisely: symmetrical). If, however,
one proceeds further and constructs the Hamiltonian of the
hydrogen atom out of $r$, $\vartheta$, $\varphi$, $\hat p_r$,
$\hat p_\vartheta$, $\hat p_\varphi$, then one obtains an operator
whose spectrum differs from that known from quantum mechanics and
is, of course, in contradiction with experiment. One is,
therefore, {\em compelled} to quantize in Cartesian
coordinates\footnote{"This assumption [of replacing classical
canonical coordinates by corresponding operators] is found in
practice to be successful only when applied with the dynamical
coordinates and momenta referring to a Cartesian system of axes
and not to more general curvilinear coordinates." \cite{Dirac47}},
transformation to curvilinear coordinates being allowed only
afterwards. In Euclidean space this restriction does not present
any problem since in this space the set of Cartesian coordinates
is preferred (it can be selected unambiguously and equivalence
within the set is ensured). Since, however, in general
pseudo-Riemannian space-time no preferred coordinates of this type
exist, this manifold seems to present an unfriendly environment
for canonical quantization.
 
As we have seen, Fermi-coordinates attached to a time-like world
line are preferred in the same sense as Cartesian coordinates in
Euclidean space so it is geometrically meaningful to prescribe
that quantization be performed in terms of them. If, in addition,
they are considered as Minkowski-coordinates in a
pseudo-Euclidean parametric space
canonical quantization becomes unambiguously defined. So far this
has been established only for parametric spaces attached to
geodesics
but in the next section this point of view will be extended to a
class of accelerating bodies of reference.
 
\section{The two types of coordinates. Accelerating bodies
of reference.}
 
Let us consider the case when the body of reference under
consideration is accelerating, i.e. moves on a general time-like
world line $\cal L$ which is not geodesic. The tetrad field along
$\cal L$ is obtained by Fermi -Walker transport rather than
parallel transport. The same kind of reasoning which for a
geodesic
$\cal G$ leads to the relations $g_{ij}({\cal G})=\eta_{ij}$,
$\Gamma_{jk}^i({\cal G})=0$ permits us to determine $g_{ij}({\cal
L})$ and $\Gamma_{jk}^i({\cal L})$ --- or
$\displaystyle \left (\frac{\partial
g_{ij}}{\partial x^k}\right )_{\cal L}$ --- again
(see the Appendix). Let us confine ourselves
to the most important case of constant acceleration when in
Fermi-coordinates $w^i=constant$ along $\cal L$. Then, up to first
order in Fermi coordinates $x,y,z$ the metric tensor is given by
the relation
\[g_{ij} = \eta_{ij} + \frac{2}{c^2}(\vec w\cdot\vec
r)\cdot\delta_{i0}\delta_{j0},\]
where $x^0,x^1,x^2,x^3\equiv ct,x,y,z$ and $\vec r = (x,y,z)$.
For a laboratory at rest on the Earth this formula
gives $g_{ij} = \eta_{ij}$ except for $\displaystyle
g_{00}=1+\frac{2\Phi}{c^2}$ where $\Phi = gz$ is the gravitational
potential in the vicinity of the Erth's surface.
 
Parametric coordinates are numerically equal to the Fermi
coordinates and in a neighbourhood {\sf N} of
$\cal L$ even their metric properties coincide in these
coordinates. This coincidence ensures the applicability of the
results of parametric space quantum theoretical calculations to
space-time within local frames. On the other hand, as we have
stressed in the preceding section, the structure of the parametric
space must be pseudo-Euclidean. Therefore, in Fermi coordinates
the fundamental quadratic form $ds^2 = F(z)\cdot
c^2dt^2-dx^2-dy^2-dz^2$ in this space must be such, that the
Riemann-tensor
vanish and at $z=0$ the relations $F=1$ and $\displaystyle
\frac{dF}{dz}=\frac{2g}{c^2}$ fulfill. These conditions have the
unique solution $\displaystyle F(z) = \left
(1+\frac{gz}{c^2}\right )^2$ (see the Appendix). The metric in
$\displaystyle ds^2 = \left (1+\frac{gz}{c^2}\right
)^2c^2dt^2-dz^2$ is identical to the Rindler
metric \cite{Rindler66} with
the coordinate singularity shifted from $z=0$ to $z= -c^2/g$. This
quadratic form is transformed to Minkowskian (primed) coordinates
by the formulae
\begin{equation}
\begin{gathered}
ct' = \frac{c^2}{g}\left (1+\frac{g}{c^2}z\right
)\sinh\frac{gt}{c}\\
z' = \frac{c^2}{g}\left [\left (1+\frac{g}{c^2}z\right
)\cosh\frac{gt}{c}-1\right ],
\end{gathered}\label{eq:b}
\end{equation}
which must be supplemented by $x'=x$, $y'=y$.
 
When the presence of gravitation cannot be neglected quantization
must be
performed in these primed coordinates. An investigation of this
kind is the COW experiment \cite{Collela75} in which the effect of
the weight
of the neutron on the interference pattern was studied. In primed
coordinates the Schr\"{o}dinger equation is the one for a free
neutron:
\begin{equation}
i\hbar\frac{\partial\psi '}{\partial t'}=-\frac{\hbar^2}{2m}\left
(
\frac{\partial^2\psi '}{\partial {x'}^2} +
\frac{\partial^2\psi '}{\partial {y'}^2} +
\frac{\partial^2\psi '}{\partial {z'}^2}\right ).\label{eq: c}
\end{equation}
Since the observations in space-time are expressed in terms of
the (unprimed) Fermi-coordinates it is
expedient to transform this equation to unprimed coordinates. In
this nonrelativistic case the $c\longrightarrow\infty$ form of
(\ref{eq:b}) is relevant:
\[t'=t,\qquad x'=x,\qquad y'=y,\qquad z'=\frac{1}{2}gt^2.\]
Then, introducing $\psi (x,y,z,t)$ instead of $\psi'(x',y',z',t')$
by the relation
\begin{equation}
\psi=\psi '\cdot e^{\displaystyle -ig\left(\frac{m}{\hbar}tz +
\frac{mg}{6\hbar}t^3\right )}\label{eq: d}
\end{equation}
we transform (\ref{eq: c}) to
\[i\hbar\frac{\partial\psi}{\partial
t}=-\frac{\hbar^2}{2m}\left (
\frac{\partial^2\psi}{\partial x^2} +
\frac{\partial^2\psi}{\partial y^2} +
\frac{\partial^2\psi}{\partial z^2}\right )+mgz\cdot\psi.\]
The effect of the last term on the interference pattern was indeed
observed.
 
Consider now the Pound-Rebka experiment \cite{Pound60} in which the
existence
of the gravitational red shift was demonstrated in the laboratory.
>From the principal point of view two distinct laboratories were
involved in the experiment, one above the other and both at
rest in the gravitational
field of the Earth. As we have already emphasized the red shift
itself is a purely classical phenomenon which is contained in the
propagation matrix between the two laboratories. The role
of quantum electrodynamics is limited to the treatment of the
emission in the lower and absorption in the upper laboratory (or
vice versa) both of which were M\"{o}ssbauer-transitions. Though
the influence of terrestrial gravitation on these processes is
certainly negligible it is a principal question how to calculate
them in our local approach.
 
The first step is to formulate quantum field theory in the
parameter space attached to either laboratory as body of
reference, accelerating
upward with constant acceleration. Just as in the case of the
Schr\"{o}dinger-equation discussed above the field quantization
also has to be performed in the primed Minkowski-coordinates. In
particular, the ground state must be identified with the vacuum in
the primed description (primed vacuum). Having this done, the
transcription to unprimed Fermi-coordinates must follow with the
aid of an appropriate unitary transformation --- the analogue of
(\ref{eq: d}), --- since the physical conditions are given in
terms
of the unprimed coordinates: the proper time is $t$ rather than
$t'$ and the radiating and absorbing atom is at rest with respect
to the coordinates $x,y,z$ rather than $x',y',z'$.
 
For a scalar field the unitary transformation involved here has
been dealt with earlier in great detail (Ref.1. Section 4.5). The
motivation for this
study was to illustrate the peculiarities of field quantization in
curved space-time on an example which allowed the comparison of
quantization in Minkowski and curvilinear coordinates. The unitary
transformation was found to have the form of a
Bogoliubov-transformation of the primed emission and absorption
operators. The primed vacuum which is the ground state in our
approach contains particles which can be absorbed by the unprimed
absorption operators. As a result, the detectors which are at rest
in the gravitational field of the Earth are expected to
continuously emit particles, though the rate of this effect is
very low.
 
The mathematics behind this phenomenon is precisely the same which
led to the recognition of the Unruh-effect, the radiation of
detectors accelerated in the space-time of special relativity. The
context is, however, quite different. First, the scene is
parametric space rather than space-time. Second, the approach
outlined here leads to the radiation of detectors which are at
rest in the static gravitational field of the Earth, i.e.
accelerating in the sense of performing non-geodesic motion. When,
on the other hand, this same situation is described in the global
approach, choosing a procedure based on
Schwarzschild-time (static quantization), then one obtains that
detectors are nonradiating when at rest in
Schwarzschild-coordinates --- but in general radiate when
moving on geodesics (i.e. at rest in local inertial frames). We
note finally that the Unruh-effect follows from our approach too.
 
\section{Compatibility with EPR}
 
Our local approach does not introduce any change into the
structure
of quantum theory. In particular, it preserves all its nonlocal
features. These properties, however, may appear {\em distorted}
when phenomena extended to very large domains of space-time,
larger than {\sf N}, are cosidered.
 
The fundamental nonlocal nature of quantum mechanics is most
clearly expressed in the EPR-type correlations violating
Bell-inequalities \cite{Bell65, Clauser78} . The observation of
this phenomenon requires
three macroscopic instruments: the source of the correlated pairs
of --- say --- photons and the two detectors which measure their
directions of polarization \cite{Aspect81} . These instruments may
belong
to different bodies of reference, locating at arbitrarily large
--- even cosmic --- distances from each other. The natural
question arises: what the
local approach can tell us on the correlations under such unusual
circumstances?
 
For the description of the phenomenon let us choose the parameter
space attached to the source of the photon pairs. Then the
description of the source and the emitted pair will be undisturbed
by
the metrical incongruence of the parameter space and space-time.
On the other hand, the detectors may be located in such domains of
space-time, where this incongruence is already
significant\footnote{We confine our discussion to that part of the
space-time which is covered by the Fermi-coordinates of the
world line of the source. Questions, concerning the extension
of this domain, are left for future study.} and,
therefore, the properties of the photons may turned out distorted
as described in the parameter space of the source.
 
As mentioned earlier, the trajectories of the photons will deviate
the more from the paths given by the parametric space
quasiclassical
calculation the farther the detectors are from the source.
The helicities, however, being invariant on the
geodesics in both space-time and parameter space, do not
experience distortions at all. Therefore, as far as the helicities
are concerned, calculations performed in the parameter space of
the source lead to reliable conclusions even for faraway
detectors. Since quantum theory in
the parameter space of a given body of reference is just
that quantum theory which we know from textbooks, the distance
Bell-inequatlities remain violated even on cosmic scale.
This is a bit disappointing
conclusion since the local approach might have expected to give
different
results for EPR type experiments within terrestrial laboratory
on the one side and on cosmic
scale on the other which could --- at least in principle --- to
distinguish between the global and local approaches.
However, the version of the theory, pursued in the present work,
seems not to offer us this opportunity.
 
\section{Summary}
 
Our accepted manner to describe physical reality does not assign
decisive role to the perspective --- or standpoint --- of the
observer who is the subject of the description. This is true even
when the world does not seem the same from different points of
observation as e.g. in the realm of the cosmological principle.
For example, the Solar System can be correctly described from
both the standpoint of the Sun and the Earth since the superiority
of the heliocentric system consists certainly not in the
reliability of the former and unreliability of the latter
description.
 
As a matter of fact, the version of the local approach
outlined in the present work offers an
alternative role to the point of observation as far as the
phenomena of quantum physics are concerned. The {\em possibility}
of description remains independent of the standpoint but its {\em
validity} becomes depending on it in a significant way. But since
observation can in principle be performed from any macroscopic
body of reference no sensible questions remain necessarily
unanswered by this narrowing of the horizon. If they do the local
approach must be abandoned.
 
In the limit of flat space-time our local approach becomes
identical with the usual space-time form of quantum theory, since
in this case it is possible to identify the parameter spaces
with the whole of the space-time. In this limit viewing the world
from different bodies of reference gives equally valid
descriptions which differ from each other solely by the choice
of the coordinate systems.

\appendix
\section{Appendix}
 
The Fermi-Walker transport of a vector $V^i$ along a time-like
world line $\cal L$ is given by the formula
\begin{gather}
\frac{DV^i}{d\tau} =
\frac{1}{c^2}g_{kl}(w^iu^k-u^iw^k)V^l,\tag{A1}
\end{gather}
in which $u^i$ is the four-velocity ($u\cdot u=c^2$),
$\displaystyle w^i=\frac{Du^i}{d\tau}$ is the four-acceleration
and $\displaystyle\frac{D}{d\tau}$ the invariant derivative along
$\cal L$:
\begin{gather}
\frac{DV^i}{d\tau}=\frac{dV^i}{d\tau}+\Gamma_{kj}^iV^ju^k.\tag{A2}
\end{gather}
 
The defining properties of the Fermi-Walker transport are:
\begin{enumerate}
\item the tangent vectors are transported into each-other,
\item the scalar product is left invariant and
\item for a geodesic $(w^i=0)$ it reduces to parallel transport
(absence of rotation). \end{enumerate}
 
Let us choose Fermi-coordinates (Section 2) attached to $\cal L$
and consider the space-like geodesics, connecting $E$ and $P$,
given by the formulae
\begin{gather}
t=constant,\qquad x=l\cdot n^x,\qquad y=l\cdot n^y,\qquad z=l\cdot
n^z,\notag\\
n^x=\cos\vartheta\cos\varphi ,\qquad
n^y=\cos\vartheta\sin\varphi ,\qquad
n^z=\sin\vartheta ,\notag
\end{gather}
$\vartheta$ and $\varphi$ being the polar angles of the geodesic
at $P$. These expressions satisfy the geodesic equation
\[\frac{d^2x^i}{dl^2} +
\Gamma_{jk}^i\frac{dx^j}{dl}\frac{dx^k}{dl}=0,\]
which leads to
$\Gamma_{\mu\nu}^in^\mu n^\nu = 0$
($i = ct,x,y,z;\;\mu ,\nu = x,y,z$).
Since at $P$ $n^\mu$ may point in any 3-direction, we obtain for a
symmetric connection
\[\Gamma_{\mu\nu}^i({\cal L})=0.\]
 
Consider now the elements of the tetrad field along $\cal L$ whose
components are $e_{(m)}^i=\delta_m^i$. According to the definition
of the Fermi-coordinates they satisfy (A1):
\begin{gather}
\Gamma_{kj}^i\cdot\delta_m^ju^k =
\frac{1}{c^2}g_{kl}(w^iu^k-u^iw^k)\delta_m^l.\tag{A3}
\end{gather}
The construction of the Fermi-coordinates also ensure that along
the coordinate lines which cross at $P$ the coordinates are
measured by the length (or proper time) on them. From this
and from the pseudo-orthonormality of the tetrads it follows that
\begin{gather}
g_{ij}({\cal L})=\eta_{ij}.\tag{A4}
\end{gather}
Since, moreover, $u^i=c\cdot\delta_0^i$, (A3) can be transformed
into
\[\Gamma_{0m}^i=\frac{1}{c^2}(w^ig_{0m}-\delta_0^iw_m),\]
from which we obtain that the only nonzero components of the
Christoffel-symbol are
\begin{gather}
\Gamma_{00}^\mu({\cal L})=\Gamma_{\mu 0}^0({\cal L}) =
\frac{1}{c^2}w^\mu.\tag{A5}
\end{gather}
Notice that for a geodesic (A4) and (A5) reduces to
\[g_{ij}({\cal G})=\eta_{ij},\qquad \Gamma_{jk}^i({\cal G})=0.\]
For a general world line we have up to first order in $x,y,z$
\begin{gather}
g_{ij} = \eta_{ij} + a_{ij}x + b_{ij}y + c_{ij}z.\tag{A6}
\end{gather}
If $w^i=constant$ the coefficients in (A6) are also constants.
 
Choose $k=1$ ($x^1\equiv x$) in the formula
\[\frac{\partial g_{ij}}{\partial x^k}-\Gamma_{ik}^lg_{lj} -
\Gamma_{jk}^lg_{il}=0\]
which expresses the vanishing of the covariant derivative of the
metric tensor and substitute (A6) into it. We obtain
\[a_{ij} = \Gamma_{ix}^l\eta_{lj} + \Gamma_{jx}^l\eta_{il}.\]
Then from (A5) it follows that the only nonzero component of
$a_{ij}$ is $\displaystyle a_{00}=\frac{2}{c^2}w^x$.
Analogous result may be obtained for $b_{ij}$ and $c_{ij}$ too, so
we find that up to first order in $\vec r = (x,y,z)$
\[g_{ij}=\eta_{ij}+\frac{2}{c^2}(\vec w\cdot\vec
r)\cdot\delta_{i0}\cdot\delta_{j0},\]
as indicated in Section 3.
 
\vspace{3mm}
 
\centerline{* * *}
 
\vspace{3mm}
 
Consider the quadratic form
\[ds^2 = F(z)\cdot c^2dt^2 - dx^2 - dy^2 - dz^2,\]
and try to choose $F(z)$ so as to make the Riemann-tensor to
vanish.
 
The curvature 2-form ${\cal R}_{.j}^i$ has a single independent
nonzero component, say
\[{\cal R}_{.0}^z = \left (\frac{{F'}^2}{4F} -
\frac{1}{2}F''\right )d(ct)\wedge dz.\]
The function $F(z)$ must, therefore, satisfy the equation
\[F'' - \frac{{F'}^2}{2F}=0,\]
the general solution of which is $F(z)=(a+bz)^2$.

\end{document}